\documentclass[aps,prl,twocolumn,showpacs,superscriptaddress,floatfix]{revtex4-2}

\usepackage{amsmath,amssymb}
\usepackage{graphicx}
\usepackage{dcolumn}
\usepackage{bm}
\usepackage{booktabs}
\usepackage[utf8]{inputenc}
\usepackage{microtype}
\usepackage{xcolor}

\begin{document}

\title{Response to: ``Isotropic deceleration and near-zero baseline acceleration in Pantheon+ supernovae: new arguments in the dark energy debate''}

\author{Animesh Sah}
\affiliation{Department of High Energy Physics, Tata Institute of Fundamental Research, Homi Bhabha Road, Mumbai 400005, India}

\author{Mohamed Rameez}
\affiliation{Department of High Energy Physics, Tata Institute of Fundamental Research, Homi Bhabha Road, Mumbai 400005, India}

\author{Subir Sarkar}
\affiliation{Rudolf Peierls Centre for Theoretical Physics, University of Oxford, Parks Road, Oxford OX1 3PU, United Kingdom}

\date{\today}

\begin{abstract}
Ray \emph{et al.} \cite{Ray:2026hdi} have claimed that the acceleration of the Hubble expansion rate inferred from Type Ia supernovae in the Pantheon+ catalogue ``is equally strong in the CMB dipole ($q_m = + 1.45$) and anti-dipole hemispheres ($+ 1.55$), directly contradicting the SRS26 anisotropy argument''. Here SRS26 refers to our analysis \cite{Sah:2026mgk} which showed that ``locally $q_0$ has a strong dipole anisotropy aligned approximately with the bulk flow, and only a small monopole component remains at distances exceeding a few hundred Mpc''. We find  that these authors confused Equatorial with Galactic coordinates, thereby getting the CMB dipole direction wrong. Hence their conclusion is baseless.

\end{abstract}


\maketitle

A significant dipole anisotropy approximately aligned with the direction of the CMB dipole, was reported \cite{Sah:2026mgk} in the deceleration parameter $q_0$ inferred from Type Ia supernovae (SNe~Ia) in the Pantheon+ catalogue, as is expected for a `tilted observer' in a bulk flow~\cite{Sah:2024csa}. Moreover the isotropic component of $q_0$ was found to be \emph{positive} after correcting the supernova magnitudes for a recently identified correlation with their progenitor age \cite{Chung:2025cgv,Chung:2026jhm}. These observations, both separately and together, argue against cosmic acceleration driven by dark energy.

Ray \emph{et al.} \cite{Ray:2026hdi} (R26 ) 
divide the Pantheon+ sample into hemispheres at $\cos\theta = 0$, where $\theta$ is the angle between each SNe~Ia's sky position and a reference direction stated to be the CMB dipole direction --- which is ($\ell, b) = (264^\circ, 48^\circ$) in Galactic coordinates. Their Fig.~3 caption reports $N=724$ supernovae with $\cos\theta \geq 0$ and $N=840$ with $\cos\theta < 0$, for a sample of $N=1564$ supernovae in the redshift range $0.00937 < z_{\rm hel} \leq 0.8$.

The Pantheon+SH0ES data release \cite{Scolnic:2021amr} however tabulates supernova sky positions in equatorial coordinates (ICRS), i.e. Right Ascension ($\alpha$) and Declination ($\delta$). To compute $\cos\theta$ between a supernova and a reference direction, they need to be expressed in the same coordinates:
\begin{equation}
\cos\theta = \sin\delta \sin\delta_0 
 + \cos\delta \cos\delta_0 \cos(\alpha - \alpha_0),
\label{eq:costheta}
\end{equation}
hence the reference direction $\alpha_0, \delta_0$ should also be in \emph{
equatorial} coordinates. The CMB dipole direction, transformed to ICRS, is in fact:
($\alpha_0, \delta_0) = (167.8^\circ, -7.1^\circ$).

We reproduce the sample selection of R26 ($N=1564$) and evaluate Eq.(~\ref{eq:costheta}) for two choices of $(\alpha_0, \delta_0)$:
\begin{enumerate}
\item the values ($264^\circ, 48^\circ$) used as equatorial coordinates, \emph{without} transforming from Galactic to ICRS, and 
\item the correctly transformed CMB dipole direction:
($167.8^\circ, -7.1^\circ$) in equatorial coordinates.
\end{enumerate}

The first row of Table~\ref{tab:hem_split} reproduces  the $N=724/840$ split reported in the caption of Fig.~3 of R26. 
This shows that R26 erred in evaluating Eq.(~\ref{eq:costheta}) taking the numerical values $(264^\circ, 48^\circ)$ to be $(\alpha_0, \delta_0)$, i.e. \emph{without} applying the Galactic-to-ICRS transformation required by their stated coordinate labels $(\ell, b)$. 
The second row of Table~\ref{tab:hem_split} shows that the correctly transformed direction yields a substantially different partition of the sample.
We conclude that the hemisphere test in Fig.~3 of R26 is not in fact centred on the CMB dipole direction. 

Indeed if we choose the direction ($264^\circ,48^\circ$) specified by R26 to be in equatorial coordinates, and fit a dipole in this direction, 
we find
$\Delta\mathrm{LLH} \approx 0.2$, i.e. no evidence for one. This is as expected since $(264^\circ,48^\circ)$ in equatorial coordinates is not the CMB dipole direction.

 It is evident that Ray \emph{et al.} \cite{Ray:2026hdi} made a technical error, hence their assertion concerning Sah \emph{et al.}  \cite{Sah:2026mgk} is invalid.

\begin{table}[h]
\centering
\caption{Hemisphere split of the $N=1564$ Pantheon+SH0ES sample
under 2 interpretations of the R26 reference direction}
\label{tab:hem_split}
\begin{tabular}{lcc}
\hline
Reference direction [ICRS] & $N(\cos\theta \geq 0)$ & $N(\cos\theta < 0)$ \\
\hline
$(264^\circ, 48^\circ)$ taken to be $(\alpha_0,\delta_0)$ & 724 & 840 \\
CMB dipole $(167.8^\circ, -7.1^\circ)$ & 539 & 1025 \\
\hline
\end{tabular}
\end{table}




\end{document}